\documentclass[%
 reprint,
superscriptaddress,
 amsmath,amssymb,
 aps
]{revtex4-2}

\usepackage{graphicx}
\usepackage{dcolumn}
\usepackage{bm}
\DeclareMathAlphabet{\mathpzc}{OT1}{pzc}{m}{it}
\usepackage{lineno}

\usepackage[caption=false]{subfig}
\usepackage{amssymb}
\usepackage{amsmath}
\usepackage{graphicx,bm}
\usepackage{verbatim}
\usepackage{xcolor, soul}
\sethlcolor{yellow}

\usepackage{float}

\begin{document}


\title{Efficient weighted-ensemble network simulations of the SIS model of epidemics}
\author{Elad Korngut\textsuperscript{1}\thanks{elad.korngut@mail.huji.ac.il}, Ohad Vilk\textsuperscript{1,2} and Michael Assaf\thanks{michael.assaf@mail.huji.ac.il}}
\affiliation{Racah Institute of Physics, Hebrew University of Jerusalem, Jerusalem 91904, Israel \\
\textsuperscript{2}Movement Ecology Lab, Department of Ecology, Evolution and Behavior, Alexander Silberman Institute of Life Sciences, Faculty of Science, The Hebrew University of Jerusalem, Jerusalem 91904, Israel}

\begin{abstract}
The presence of erratic or unstable paths in standard kinetic Monte Carlo simulations significantly undermines the accurate simulation and sampling of transition pathways. While typically reliable methods, such as the Gillespie algorithm, are employed to simulate such paths, they encounter challenges in efficiently identifying rare events due to their sequential nature and reliance on exact Monte Carlo sampling. In contrast, the weighted ensemble method effectively samples rare events and accelerates the exploration of complex reaction pathways by distributing computational resources among multiple replicas, where each replica is assigned a weight reflecting its importance, and evolves independently from the others. Here, we implement the highly efficient and robust weighted ensemble method to model susceptible-infected-susceptible (SIS) dynamics on large heterogeneous population networks, and explore the interplay between stochasticity and contact heterogeneity which ultimately gives rise to disease clearance. {\color{black}Studying a wide variety of networks characterized by fat-tailed asymmetric degree distributions, we are able to compute the mean time to extinction and quasi-stationary distribution around it} in previously-inaccessible parameter regimes. 
\end{abstract}

\maketitle

\section{\label{sec:Intro}Introduction}
Compartmental models are widely used in epidemiology for understanding the dynamics of infectious diseases within populations, aiding in the analysis of transmission patterns, disease prevalence, and the effectiveness of intervention strategies 
\cite{background_inf_models,Vespignani_scale_free,vespignani_sis_formulation,Vespignani_Pastor_epidemic_scale_free,keeling2005networks,Dorogovtsev_percolation,keeling2011modeling,anneald_net}. The crux of these models is to provide a structured framework to analyze complex interactions between various compartments representing different disease states, such as susceptible and infected individuals. Recently, compartmental models have played a key role in modeling the transmission and mitigating the COVID-19 pandemic \cite{vespignani_covid_travel_restrictions,vespignani_compartment_covid19}. 
Apart from epidemiology, such models have broad applications, e.g., in reliability engineering, election result forecasting, and the spread of computer viruses~\cite{electric_network,elections,computer_virus}. One of the simplest compartmental models is the susceptible-infected-susceptible (SIS) model, which is widely used for diseases in which the endemic state persists  for an extended period, see e.g.~\cite{Vespignani_scale_free,vespignani_sis_formulation,keeling2005networks,Dorogovtsev_percolation,keeling2011modeling,
moore2000epidemics,karrer2010message,
Pastor-Satorras_sis_backtracking}. 
Amongst others, the SIS model effectively describes the dynamics of diseases such as influenza, the common cold, tuberculosis, and sexually transmitted infections like chlamydia and syphilis~\cite{keeling2005networks}.

In the SIS model, a population is divided into two distinct groups, where individuals are either susceptible (S) to the infection or currently infected (I). Infected individuals may recover and revert to a susceptible state through treatment or natural recovery, while susceptible individuals can contract the disease upon contact with infected individuals. In the limit of very large populations, as long as the infection rate  exceeds the recovery rate (see below) the model exhibits a stable endemic solution, in which the disease persists forever within the population. However, for finite populations demographic stochasticity ultimately drives the system from the stable endemic state to the unstable state, corresponding to disease extinction, via a rare large fluctuation~\cite{background_inf_models,anneald_net,NOLD1980,ovaskainen_2001,pre_mickey_2010,wkb_miki}.  Notably, despite their key role in determining the clearance probability, these large fluctuations are composed of a large number of sequential recovery events, which are challenging to detect using standard numerical methods.  

In simple one-dimensional scenarios,  computing the probabilities of such rare events can be done, e.g., by employing a semi-classical approximation on the pertinent master equations~\cite{dykman_wkb_exp,dykman2008disease}. These  scenarios include a well-mixed, or \textit{homogeneous} setting, where each individual interacts with an equal number of neighbors~\cite{ovaskainen_2001,pre_mickey_2010,wkb_miki}. Recently, more complex, \textit{heterogeneous} scenarios were analytically studied; yet, rigorous analysis is limited to regimes where heterogeneity is weak or specific~\cite{pmid29476196,clancy_same_paper,miki-jason.123.068301, korngut}, or close to the bifurcation limit of the SIS model~\cite{jason-ira-paths}. 

To overcome these challenges, numerical methods such as kinetic Monte Carlo (KMC) simulations are often used to study rare events~\cite{anneald_net,GILLESPIE1976_second_paper,doi:gillespie,gillespie_review,efficent_gillespie}.
The KMC method employs stochastic sampling to model the time evolution of systems characterized by discrete events and transitions \cite{GILLESPIE1976_second_paper,doi:gillespie}. {\color{black} {Previously, KMC algorithms have been successfully used  
to explore mean-field dynamics and epidemic thresholds in very large population networks up to $10^8$ nodes~\cite{Pastor-Satorras_correlated_networks,Eguıluz_2002_Epidemic_Threshold_Structured_Scale_Networks,Boguna_2013_Nature_Epidemic_Threshold_in_Networks}}. Yet, numerically detecting rare events is far more challenging, as} this requires a large number of KMC realizations in order to get a good accuracy, which comes at a high computational cost~ \cite{group_meeting_paper}. KMC simulations may also encounter sampling inefficiency, thus facing challenges in effectively exploring high-dimensional phase spaces in heterogeneous environments \cite{ira_path}. 

Other numerical alternatives include numerically solving the underlying set of time-dependent master equations,which yields the probability distribution of finding a specific number of individuals in each state (susceptible or infectious)~\cite{sim_master}. However, this is feasible only in low-dimensional systems, as runtime exponentially depends on the dimensionality~\cite{sim_master,korngut}. To circumvent this problem, one can  employ a semi-classical approximation to the master equation, which yields a set of Hamilton's equations that can be solved numerically, and whose number scales with the system's dimensions~\cite{dykman_wkb_exp,pre_mickey_2010,wkb_miki}. Nonetheless, these numerical schemes are less accurate and highly sensitive to the initial conditions, and thus, are less practical in  multi-dimensional cases.

In contrast, the weighted ensemble (WE) method is both efficient and accurate in sampling rare events. By distributing computational resources among multiple replicas and adjusting weights dynamically, the WE method accelerates the exploration of rare events, making it particularly suitable for studying large deviations in dynamical systems \cite{we_paper}. Previously, this method was successfully applied for ecological, biological and chemical models with a large computational benefit over ``brute-force" methods \cite{we_bio_chem_reaction,vilk_we_ali,vilk_meerson_we_brownian}. Yet, thus far it has not been used to study long-time dynamics on population networks.

In this work we demonstrate the efficiency and computational advantages of the WE method, which is shown to be vastly superior to "brute force" KMC-like methods in studying rare events on large population networks. As such, the WE method allows investigation of population networks in parameter regimes that were previously inaccessible due to the computational resources required. 
{\color{black}As a prototypical example we implement the SIS dynamics on heterogeneous networks and show that the WE method is highly effective in computing the mean time to extinction (MTE), as well as the quasi-stationary distribution (QSD) around the long-lived endemic state.} Our analysis is performed on  a wide variety of network topologies with varying heterogeneity strengh and skewness, exhibiting fat-tailed distributions, such as observed, e.g., in
social, citation and biological networks~\cite{vespignani_sis_formulation,Vespignani_scale_free,power_distrbution_defnitions,Muchnik_social_power_law,citation_power_law,power_law_bio_sis,power_law_mickey_sent}.

The paper is organized as follows. In Sec.~II we present the theoretical model and known  results  for homogeneous and weakly-heterogeneous  networks, and close to the bifurcation limit. Sections~III and IV are dedicated to presenting our numerical algorithm and the results. Finally in Sec.~V we conclude and present  future directions.

\section{\label{sec:theory}THEORETICAL FORMULATION}
We formulate the SIS model in a topologically heterogeneous network, where nodes represent individuals of an isolated population of size $N$, each capable of being in either a susceptible ($S$) or infected ($I$) state \cite{sis_book_bailey,sis_degree-correlated_networks}. The network's topology is represented by an adjacency matrix $\textbf{A}$, in which the elements indicate links between nodes, such that $A_{ij}=1$ if nodes $i$ and $j$ are connected and $0$ otherwise. Individuals can transition from being susceptible to infected only through links,  representing potential interactions between individuals, and from infected to susceptible via spontaneous recovery. We define $\beta$ as the infection rate, attributed to each link, and $\gamma$ as the recovery rate, associated with each node. 

To allow for analytical progress, we work under the annealed network approximation~\cite{anneald_net},
i.e., a mean-field approximation over an ensemble of networks. Under the annealed network approximation one may replace the adjacency 
matrix $\textbf{A}$ with its expectation value $\langle \textbf{A}\rangle$ of an ensemble of networks~\cite{anneald_net}, which for uncorrelated networks, satisfies: $\langle A_{ij}\rangle =k_i k_j/(N\langle k\rangle)$. Here $k_i$ and $k_j$ are the degrees of nodes $i$ and $j$ while $\langle k \rangle$ is the average degree. 

To proceed, we partition nodes into groups based on their degree $k$, with each group  having rates
\begin{equation}
\label{eq:transition_rate}
    I_{k}\xrightarrow{W_{k}^{+}(\mathbf{I})} I_{k}+1 , \quad
    I_{k}\xrightarrow{W_{k}^{-}(\mathbf{I})} I_{k}-1.
\end{equation}
Here, $W_{k}^{+}(\mathbf{I})$ and $W_{k}^{-}(\mathbf{I})$ are the infection and recovery rates, respectively, and $I_{k}$ is the number of infected individuals having degree $k$. While the recovery rate reads $W_{k}^{-}(\mathbf{I})=\gamma I_{k}$, the  infection rate depends on interactions with the neighbors. 
Denoting the network degree distribution by $p(k)$, there are $N_k=Np(k)$ nodes of degree $k$, such that $\sum_k N_k=N$. Thus, the infection rate satisfies
\begin{equation} \label{rate_equation}
W_{k}^{+}(\mathbf{I})=\beta k (N_k-I_k)\sum_{k'}\frac{k'I_{k'}}{N\langle k\rangle},
\end{equation}
where $1\leq k\leq k_{\text{max}}$ and $k_{\text{max}}$ is the maximal degree. 

{\color{black}{Henceforth we consider uncorrelated random networks, where the assortativity---the tendency of high-degree nodes to be connected to high-degree nodes and vice versa---is assumed to be very low~\cite{newman_asso_prl} (see Fig.~\ref{fig1}).}  In this regime, the basic reproduction number is defined as~\footnote{{\color{black}Equation~\eqref{R0} applies for random, uncorrelated networks and when $p(k)$ is not too wide. For generic networks with given adjacency matrix \textbf{A}, one can compute $R_0$ numerically, by writing $R_0=\beta/\beta_c$, where $\beta_c=1/\lambda^{(1)}$ is the critical infection rate at the epidemic threshold, and $\lambda^{(1)}$ is the largest eigenvalue of \textbf{A}. However, this calculation also requires a large spectral gap between the largest and second-largest eigenvalues, i.e., $\lambda^{(1)}\gg \lambda^{(2)}$~\cite{anneald_net,hindes_Schwartz_2017_Epidemic_extinction_paths_in_complex_networks,Pastor-Satorras_correlated_networks}.}}
\begin{equation}\label{R0}
R_{0}=\beta\langle k^2\rangle/\left(\gamma\langle k\rangle\right),
\end{equation}
}where $\left<k^i\right>$ is the $i$-th moment of the distribution $p(k)$.
In the large-population, deterministic limit,  as long as $R_0>1$,  a nontrivial stable endemic state $I^*>0$ prevails, whereas for $R_0\leq 1$ the extinct state, $I^*=0$, becomes stable~\cite{vespignani_sis_formulation,Vespignani_scale_free}.  
Yet, disease extinction via demogrpahic noise can still occur even for $R_0>1$, as long as the population is  finite. To account for this noise,  we write a set of coupled master equations for the joint probability, $P(\mathbf{I},t)$, to find $\mathbf{I}=\left\{I_{1},...,I_{k_{\text{max}}}\right\}$ infected individuals on the different nodes. Using the infection and recovery rates, Eqs.~(\ref{eq:transition_rate}) and (\ref{rate_equation}), the equation for $P(\mathbf{I},t)$ reads:
\begin{eqnarray}
        \frac{\partial P(\mathbf{I},t)}{\partial t}&&=\sum_{k=1}^{k_{\text{max}}}\left[W_{k}^{-}(I_{k}+1)P(\mathbf{I}+\mathbf{1}_{k},t)\!-\!W_{k}^{-}(I_{k}) P(\mathbf{I},t) \right. \nonumber\\
        &&\left.+W_{k}^{+}(I_{k}\!-\!1)P(\mathbf{I}\!-\!\mathbf{1}_{k},t)-W_{k}^{+}(I_{k})P(\mathbf{I},t)\right],
    \label{eq:master}
\end{eqnarray}
where $\mathbf{I}\pm\mathbf{1}_{\!j}$ denotes an increase or decrease by 1 of $I_{j}$~\cite{dykman_wkb_exp}. 

While Eq.~\eqref{eq:master} is generally unsolvable for arbitrary $p(k)$, one can proceed analytically in the  large network limit, $N\gg 1$. Here, prior to disease extinction, the system enters a long lived metastable endemic state, which slowly decays in time while simultaneously the extinction probability grows. In this case, the MTE can be computed within exponential accuracy via the Wentzel-Kramers-Brillouin (WKB) method, which  transforms Eq.~(\ref{eq:master}) into a set of Hamilton's equations. Their solution provides the MTE up to exponential accuracy, in the form $T_{\text{ext}}\sim e^{N{\cal S}}$, where ${\cal S}$ is the action barrier to extinction~\cite{dykman_wkb_exp,dykman2008disease,pre_mickey_2010,wkb_miki,jason-ira-paths}.

While the theoretical formalism for general degree distributions can be found in~\cite{jason-ira-paths}, for completeness we provide the calculation in the simplest setting of a homogeneous population network, called a random regular network, where each node has degree $\langle k\rangle=k_0$. Here, the network's degree distribution satisfies $p(k)=\delta_{k,k_0}$, and as a result, the sum in Eq.~(\ref{eq:master}) contains a single term, and one has: $W^{+}(I) = \left(\beta k_0/N\right) I(N-I)$ and $W^{-} = \gamma I$. Using the WKB method, in the leading order in $N\gg 1$, the MTE can be shown to satisfy~\cite{ovaskainen2010stochastic,wkb_miki}:
\begin{equation}\label{actregular}
T_{\text{ext}}\sim e^{N{\cal S}},\quad\quad {\cal S}=\ln R_{0}+1/R_{0}-1,
\end{equation} 
where here $R_0=\beta k_0/\gamma$.

The action barrier has also been computed for heterogeneous networks, but only in  parameter regimes, where an additional small parameter exists, allowing to significantly reduce the dimensionality of the master equation set~(\ref{eq:master}). Such is the scenario close to bifurcation, characterized by $R_{0}-1\ll 1$, for which $\mathcal{S}$ satisfies~\cite{jason-ira-paths}
\begin{equation}
\label{eq:action_small_r}
\mathcal{S} = \frac{\left< k^2\right>^3}{2\left< k^3\right>^2}\left(R_{0}-1\right)^2 +\mathcal{O}\left(R_{0} -1\right)^3.
\end{equation}
For $R_0-1\ll 1$, Eq.~(\ref{actregular}) gives way to ${\cal S}\simeq (1/2)(R_0\!-\!1)^2$, which coincides with Eq.~(\ref{eq:action_small_r}) for a homogeneous network. 

The limit of weak heterogeneity can also be analytically treated, for arbitrary $R_0$. Here, $\mathcal{S}$ 
satisfies~\cite{miki-jason.123.068301}:
\begin{eqnarray}
\label{eq:mj_action_correction}
    \mathcal{S}&=&\mathcal{S}_{0}-f(R_{0})\epsilon^{2} +\mathcal{O}\left(\epsilon\right)^3\nonumber\\
    f(R_{0})&=&\frac{(R_{0}-1)(1-12R_{0}+3R_{0}^2)+8R_{0}^2\ln(R_{0})}{4R_{0}^3},
\end{eqnarray}
where $\mathcal{S}_0$ is the result for a homogeneous network [Eq.~(\ref{actregular})], and $\epsilon=\sigma/\langle k\rangle$ is the distribution's coefficient of variation (``strength" of the network heterogeneity), with $\sigma$ being the standard deviation of $p(k)$. Notably, Eq.~(\ref{eq:mj_action_correction}) holds for weak heterogeneity, i.e., $\epsilon\ll 1$~\cite{miki-jason.123.068301}. 

At this point, an analytical calculation of the MTE in the general heterogeneous case, is beyond reach. Thus, dealing with realistic scenarios necessitates using numerical schemes for determining the action barrier. In the following we detail the numerical algorithm we have used to implement the KMC and WE network simulations.

\section{NUMERICAL SIMULATIONS}
\vspace{-2mm}
To generate the network's topology we employ the so-called configuration model that ensures no correlations between node degrees \cite{config_model}. The random networks are characterized by their degree distribution $p(k)$, whose average and coefficient of variation, are given by $\langle k\rangle$ and $\epsilon$, respectively. However, even among networks with an identical degree distribution, the adjacency matrix $\textbf{A}$ may vary. To address this variability, we generate multiple network realizations, compute the MTE for each realization, and average the results across all networks (see below). Furthermore, in all our comparisons between different networks, 
we kept $R_{0}$ constant, ensuring that the distance to threshold remains fixed. Naturally, as $R_0$ depends on the the network topology, to maintain the same  $R_0$ across different networks, see Eq.~(\ref{R0}), we adjusted the ratio  $\beta/\gamma$ such that  $\beta/\gamma=R_{0}/\left(1+\epsilon^2\right)$. 

\begin{figure}[t]
    \includegraphics[width=0.75\linewidth]{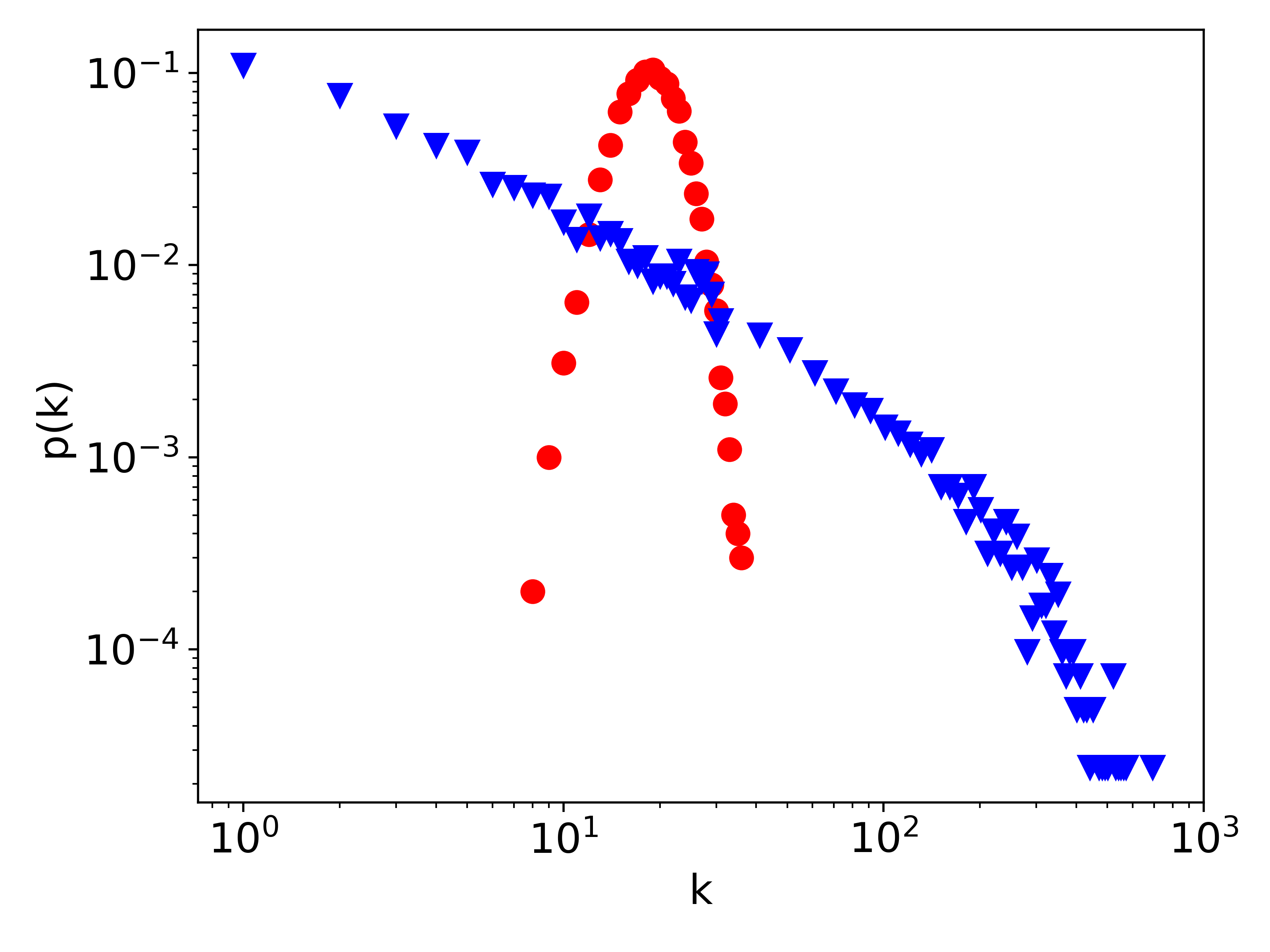}%
\vspace{-6mm}    
\caption{Log-Log plot of the gamma degree distribution. Shown is $p(k)$ versus the degree $k$, for networks of size $N=10^{4}$ with $\left< k\right>=20$ and $\epsilon=0.2$ (circles) and $\epsilon=3.0$ (triangles). {\color{black}  The assortativity of both networks is $< 10^{-3}$.}} 
    \label{fig1}
\end{figure}

To demonstrate the efficiency of our algorithm, we took four different degree distributions, based on the gamma, beta, inverse Gaussian, and log-normal distributions, see Appendix for a detailed description. In the limit of weak heterogeneity, $\epsilon\ll 1$,    these  distributions exhibit a narrow peak and low variability, with a rapidly decaying tail. Conversely, when $\epsilon={\cal O}(1)$, the  variability increases and a fat right tail  emerges. Here, due to the presence of numerous high-degree nodes, accurate representation of these distributions requires taking a large network size, see Fig.~\ref{fig1} (and also Fig.~\ref{figS1}). {\color{black}Notably, in all the simulations throughout this work, the we used random networks with negligible degree-degree correlations (see Fig.~\ref{fig1}).}

The conventional method to simulate SIS dynamics involves the KMC approach, such as the Gillespie algorithm \cite{GILLESPIE1976_second_paper,doi:gillespie,gillespie_review,efficent_gillespie}. Here, initially, a subset of nodes is randomly designated as infected, while the rest are labeled as susceptible. Transitions between states occur at exponentially distributed waiting times, where the time until the next reaction is randomly determined by the infection and recovery rates. We define the typical Gillespie time step between reactions as $\tau_{G}$, which equals the inverse of the sum of the reaction rates, and  simulate these Gillespie steps until disease clearance occurs. In the limit of large networks, the time to disease extinction follows an exponential distribution. Hence, the network's MTE and its confidence bounds are found from fitting the simulated extinction times to an exponential distribution. 
These MTEs are  averaged across several network realizations which gives rise to the overall MTE, where its error bars are determined by the standard deviation of the confidence bounds. However, this simulation method is inherently slow, since  the KMC method samples the phase space with computational time that is inversely proportional to the probability to be in a given state. This results in a significant computational effort that is not necessarily concentrated on the rare event regions. 

In contrast, the WE method involves simulating multiple realizations of the system simultaneously, each assigned weights contingent on their current state \cite{we_paper,we_simulations_compete,vilk_we_ali}. These weights are dynamically pruned to facilitate efficient exploration of the phase space by channeling computational resources towards the most relevant realizations. To determine the latter, the phase space is partitioned into bins, designating distinct regions of the system's potential states. Bins are interactively chosen (on the fly), where regions close to extinct state are set to include more instances, as detailed below (see also~\cite{code}). 
    
\begin{figure}[t]
    \includegraphics[width=0.7\linewidth]{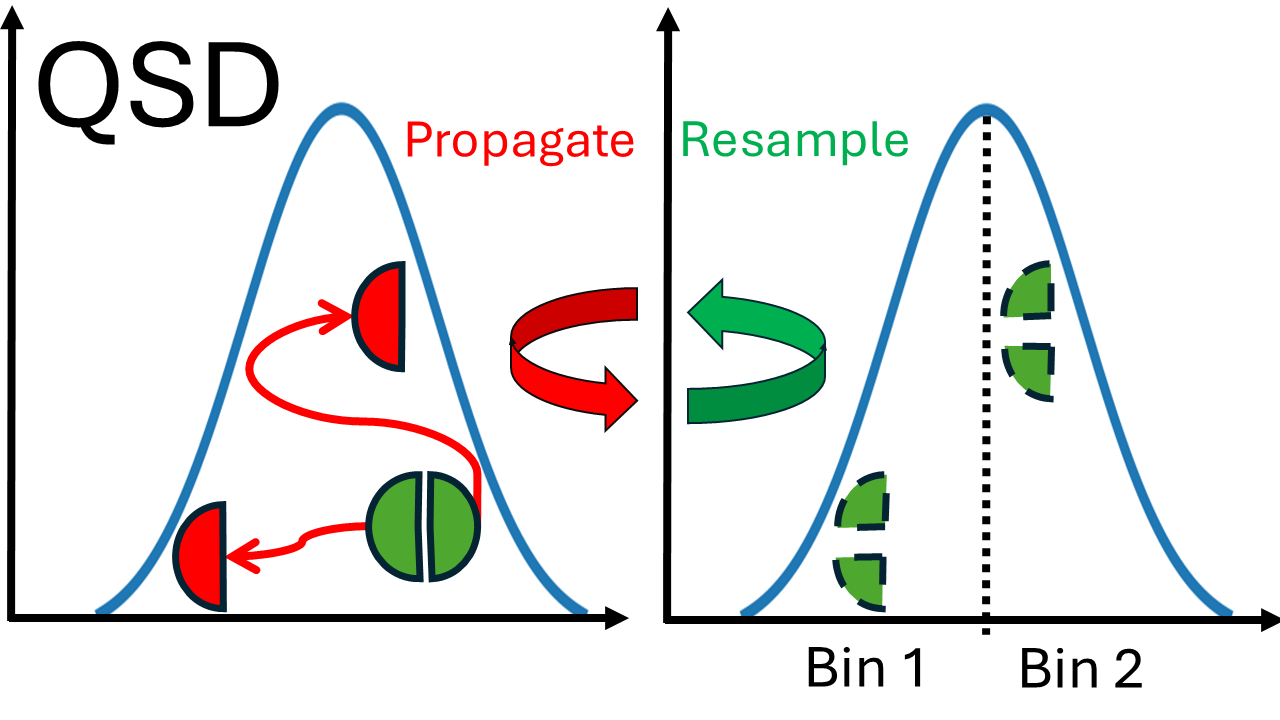}%
\vspace{-3mm}    
\caption{ {\color{black}{Illustration of a WE simulation with two realizations per bin. Left panel: Two equally-weighted realizations (green semi-circles) are generated. Using Gillespie's algorithm, both simulations are propagated up to time \(\tau_{WE}\) (red arrows) and reach new states (red semi-circles). Right panel: The resampling process includes creating new bins for realizations approaching extinction. To maintain two simulations per bin, $m=2$, realizations are then partitioned and their respective weight is halved, resulting in a weight of $1/4$ per realization. This process is iteratively repeated $M$ times, see text.}  } }
    \label{fig2}
\end{figure}

We begin by initially dividing the phase space into two bins: one for states where the overall infected density exceeds $I^*/N$ and one for those states with a lower density, where $I^*/N=1-1/R_0$ is the endemic state in a well-mixed setting, and is an adequate estimate for the  endemic state in the heterogeneous network. Within each bin, $m$ copies of a network are generated, with a fraction of randomly selected infected nodes (seeds). Each realization is assigned a weight, representing its relative importance; initially, all realizations are assigned a weight of  $1/(2m)$. Subsequently, the dynamics of each realization are simulated by Gillespie time steps until time $\tau_{WE}$, which satisfies: $\tau_{G} \ll \tau_{WE}\ll T_{\text{ext}}$. At each WE step, i.e., upon reaching time $\tau_{WE}$, 
the weight of those simulations that have undergone extinction, which determine the extinction flux, is recorded. The surviving simulations undergo a resampling process, which proceeds as follows: if a realization explores a new state with fewer infected individuals than previously recorded during each WE step, the bin closest to the extinction state is split at the new state, and the realization is replicated in the new (lowest) bin $m$ times. For the remaining bins, for those with fewer than $m$ realizations, the bin's highest-weight realizations are iteratively split, until the bin contains $m$ realizations. This splitting involves replicating the realization and dividing its weight equally. {\color{black}For bins with more than $m$ realizations, randomly chosen low-weight realizations are combined such that the realization's new weight equals the sum of the weights of all the  combined realizations, and exactly $m$ realizations remain in the bin. 
Figure~\ref{fig2} shows a simple example of the method's propagation and resampling steps.}

The resampling process ends with each bin containing $m$ weighted realizations. This WE step is performed $M$ times, and the MTE satisfies:
\begin{equation}
    T_{\text{ext}}=\frac{\tau_{WE}}{\frac{1}{M}\sum_{j=1}^{M}\phi_{j}},
    \label{eq:MTE_WE}
\end{equation}
where $\phi_{j}$ represents the extinction flux at WE step $j$.

\begin{figure}[t]
    \includegraphics[width=0.75\linewidth]{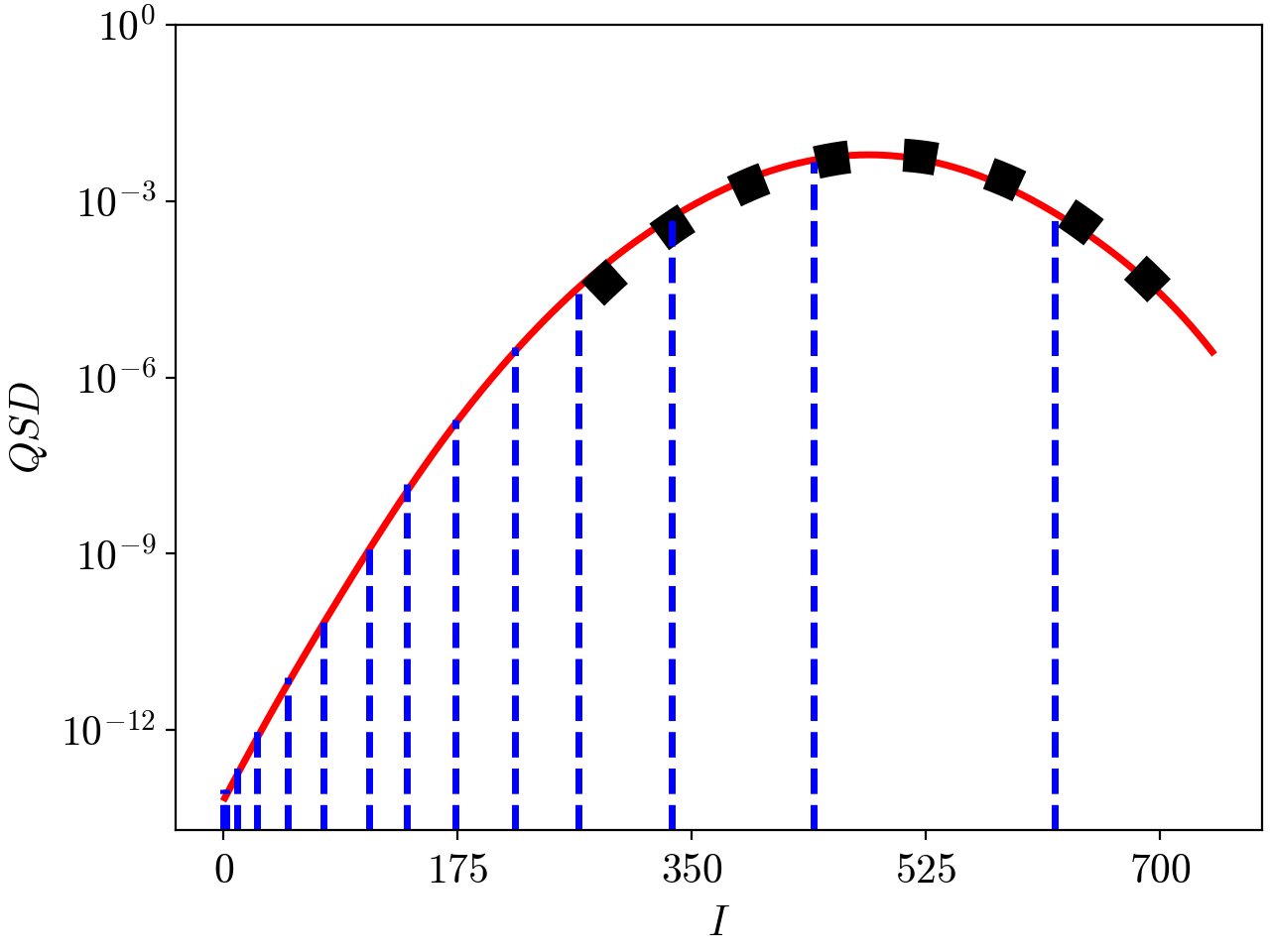}%
\vspace{-3mm}    
    \caption{{\color{black}{Semi-log plot of the QSD versus the total number of infected in the network, $I$, for a gamma network with \( N = 10^4 \) nodes, \( R_0 = 1.3 \), \( \langle k \rangle = 20 \), and \( \epsilon = 1.5 \). The solid line represents WE simulations, with parameters \( m = 10^3 \), \( \tau_{WE} = 1 \), and \( M \!=\! 70 \), while the dashed vertical lines represent the WE bin boundaries. The dotted line shows the results of a KMC simulation, which ran up to $t=10^4$ in units of $\gamma^{-1}$.
}}} 
    \label{fig3}
\end{figure}

{\color{black}In addition to \( T_{\text{ext}} \), which is found through the (inverse of the) total flux, the QSD can also be obtained by calculating the weight in each bin. For example, by dynamically constructing bins with a higher density near $I=0$, a finer resolution can be achieved in this region, allowing for accurate estimation of rare events near extinction. In contrast, KMC methods require very long realizations, in order to properly sample rare probabilities, which makes this process extremely slow and computationally expensive. Notably, by efficiently focusing on relevant regions,  the WE method achieves the same result with far fewer resources. This approach can also be readily adapted to explore any region of interest. In Fig.~\ref{fig3} we plot an example of the QSD and compare between the capabilities of the KMC and WE simulations in probing rare events.}

To estimate the error of the WE calculations of the MTE and QSD, we identify two sources of uncertainty: (i)  stochastic nature of the method, producing different results for the same network, and (ii) variability due to different network realizations.
To assess (i) we ran WE simulations with a varying number of realizations per bin $400\!<\!m\!<\!5000$, resampling time $0.2\!<\!\tau_{WE}\!<\!2$, and a fixed number of time steps $M=70$. The standard deviation of these results served as the network confidence bounds. To assess (ii) we applied the WE method on multiple network realizations. The mean over all networks provided the overall MTE, while the standard deviation of the confidence bounds gave the MTE's error bars. The latter are captured by the symbol sizes in all figures.

\begin{figure}[t]
    \includegraphics[width=0.8\linewidth]{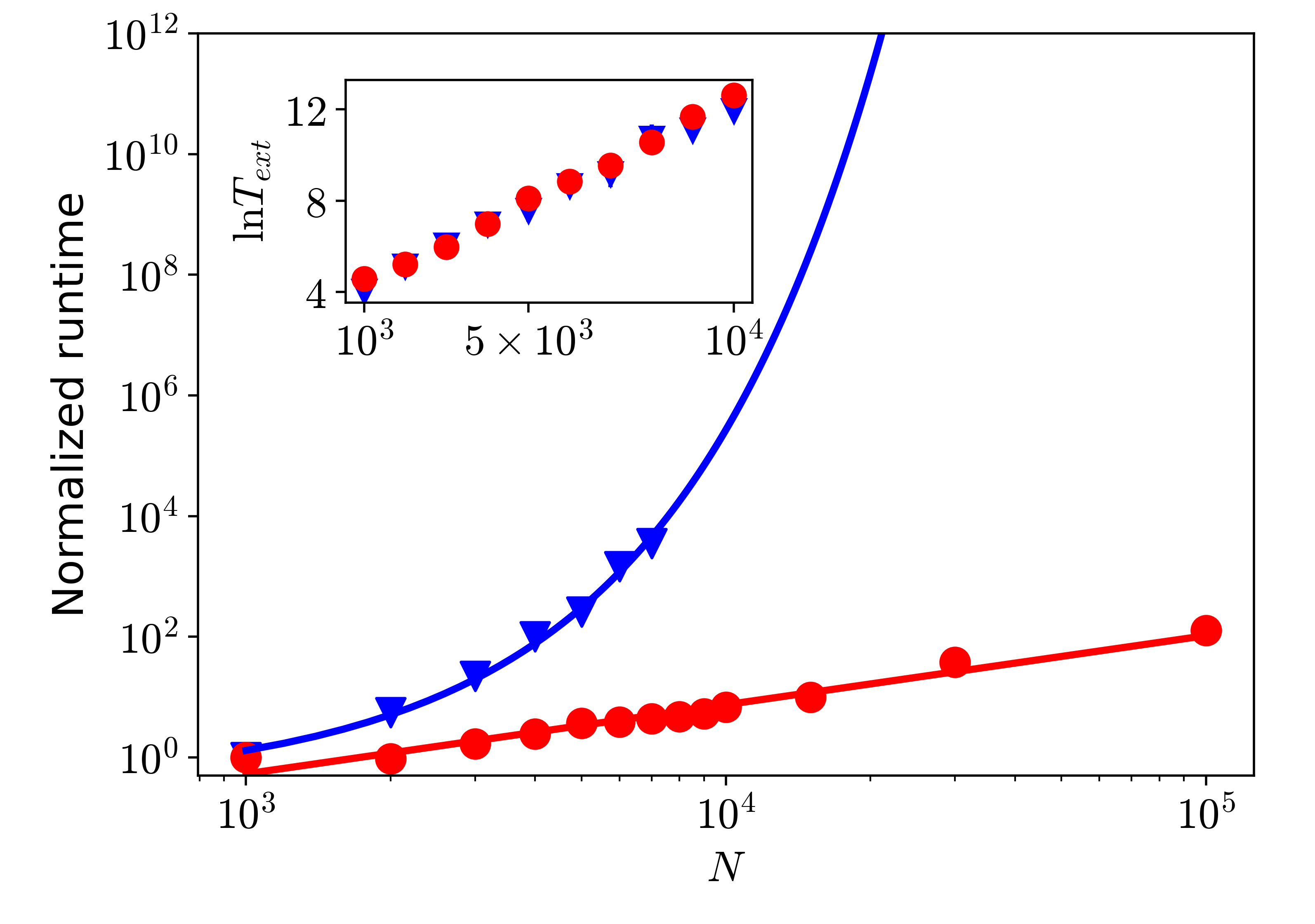}%
\vspace{-3mm}    
    \caption{A log-log plot of the runtime to obtain the MTE for each of the methods, versus the population size,  $N$, normalized by the runtime at $N=10^3$, for a gamma-distributed network. Here $\left<k\right>=10$, $\epsilon=3.0$, and $R_{0}=1.3$. {\color{black}{The WE  parameters are $m=10^3$, $M=70$ and $\tau_{WE}=1$ for $N < 10^5$, while $\tau_{WE}=0.5$ for $N=10^5$}}. Triangles and circles represent the KMC and WE methods, respectively.  The KMC scales exponentially with $N$ (blue line -- exponential fit), while the WE method scales linearly with $N$ (red line -- linear fit).
Inset shows the logarithm of the MTE versus $N$. Symbols and parameters are the same as in the main figure.} 
    \label{fig4}
\end{figure}

In Fig.~\ref{fig4} we compare the runtimes of the KMC and WE methods. 
One can see that while the KMC scales exponentially with $N$, the WE method is significantly faster, scaling linearly with $N$; e.g., if for $N=10^3$ the runtimes were equal, for $N=10^4$, the WE method out-competes the KMC by a factor of $\sim10^4$. The inset of Fig.~\ref{fig4} compares the MTE results of the two methods, confirming the WE method's accuracy.

\section{Results}
\begin{figure}[t]
    \includegraphics[width=0.8\linewidth]{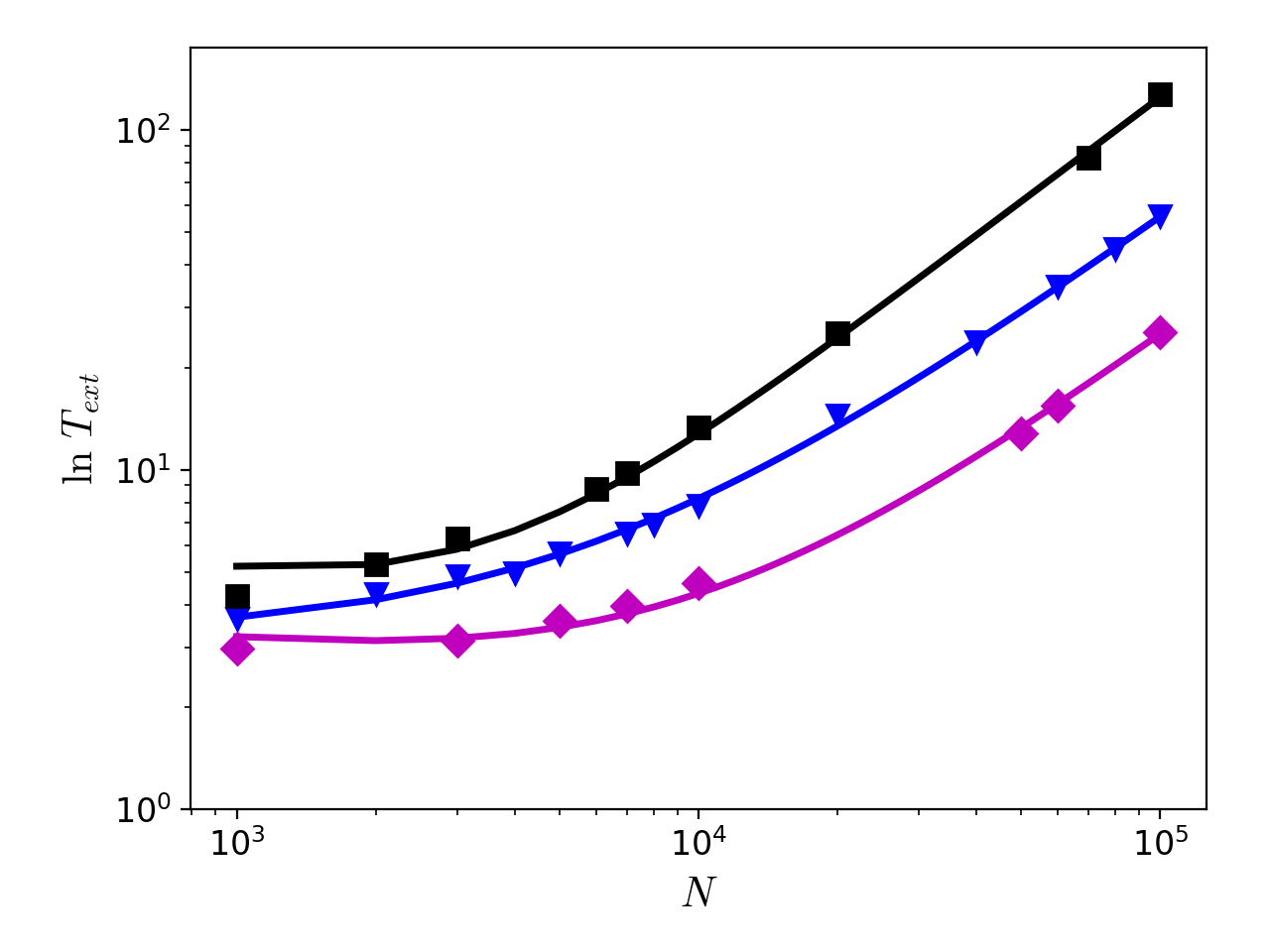}%
\vspace{-6mm}    
\caption{Log-Log plot of the logarithm of MTE versus $N$, for inverse-Gaussian (triangles), beta (squares) and log-normal (diamonds) networks. Here, $\left<k\right>=10$, $\epsilon=3.0$,  $R_{0}=1.3$, and the {\color{black}WE parameters are \( M = 70 \),  $m=500$  and \( \tau_{WE} = 1 \) for \( N < 10^{5} \), and $m=10^3$  and \( \tau_{WE} = 0.5 \) for $N=10^5$.} Solid lines are linear fits with logarithmic corrections, see text. 
} 
    \label{fig5}
\end{figure}

We employ the WE method to compute the MTE for four different networks, each with a different degree distribution. Results for the gamma distribution are shown in Fig.~\ref{fig4}, while results for the inverse-Gaussian, beta and log-normal  distributions as a function of the population size are shown in Fig.~\ref{fig5}. Based on previous work, the general form of the MTE is expected to have the following scaling with $N$: $T_{\text{ext}}\simeq A\,N^{\alpha}\,e^{N{\cal S}}$~\cite{pre_mickey_2010,jason-ira-paths}, where $\alpha$ and $A$ are some constants, and ${\cal S}$ is the action barrier. To corroborate this scaling, we have fitted the logarithm of the MTE to a linear function of $N$ with logarithmic corrections. Notably, the WE method allows simulating very large networks of size $N=10^5$ and higher, which is well beyond the capabilities of the KMC method, whose runtime diverges at such network sizes (see Fig.~\ref{fig4}). In Fig.~\ref{fig5} we demonstrate that this scaling of the MTE with $N$, derived for homogeneous and weakly-heterogeneous network, also holds for strongly heterogeneous networks, with a very high COV. Even though there is an offset between different networks (indicating a different constant $A$), as long as the mean and COV are fixed, one observes good qualitative agreement across different networks.

\begin{figure}[t]
    \includegraphics[width=0.8\linewidth]{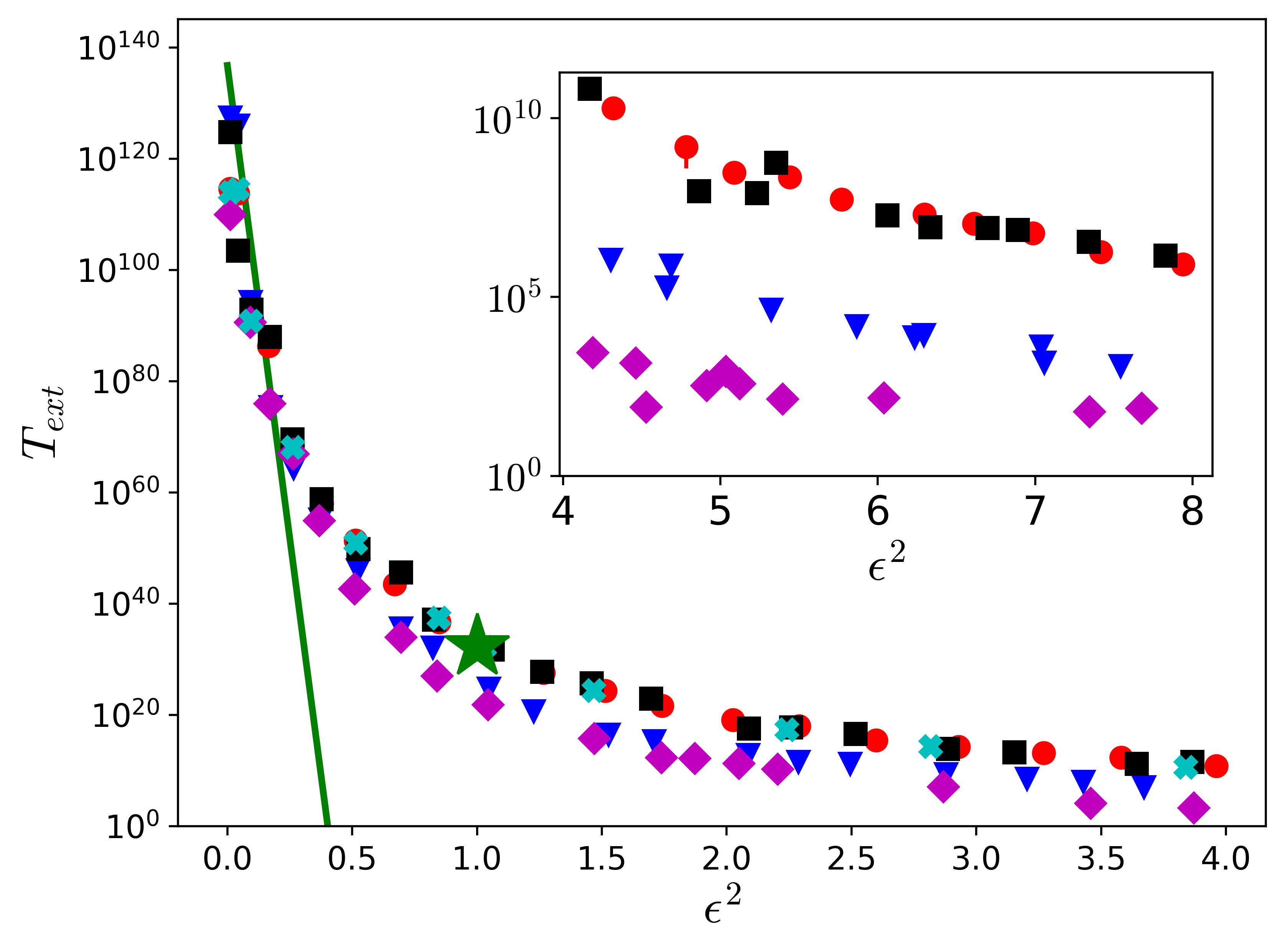}%
    \vspace{-5mm}    
    \caption{The MTE versus $\epsilon^2$, for  gamma (circles), inverse-Gaussian (triangles), beta (squares) and log-normal (diamonds) networks. Here, $N=10^4$, $\left<k\right>=20$, $R_{0}=1.3$, {\color{black}{and WE parameters are  \( M \!=\! 70 \),  \( m\!=\!10^{3} \) and \( \tau_{WE} \!=\! 1.0 \) for  simulations with \( \epsilon \!>\! 0.5 \), while $1000\!<\!m\!<\!3000$ and \( 0.2\!<\!\tau_{WE}\!<\!0.6 \) for simulations with $\epsilon\!<\!0.5$}}. The star denotes the MTE for a network with exponential degree distribution, with $\epsilon=1$, while the solid line represents the analytical prediction for weak heterogeneity [Eq.~\eqref{eq:mj_action_correction}]. Inset shows the large-$\epsilon$ regime where the results for the various networks depart.} 
    \label{fig6}
\end{figure}

The impact of the COV, $\epsilon$, on the MTE is shown in Fig.~\ref{fig6}, where the MTE is plotted against $\epsilon^2$. Notably, the WE method allows us to simulate the dynamics on large networks over a large span of COVs, which yield a huge range of MTEs, well beyond the capabilities of the KMC method. Our results in Fig.~\ref{fig6} indicate that the MTE strongly decreases with increasing $\epsilon$; namely, as heterogeneity increases, disease clearance becomes more and more likely, even for very large networks. For very strong heterogeneity ($\epsilon=\sqrt{8}$, see inset), the MTE drops drastically. Here, apparently the action barrier multiplying $N$ in the exponent, see Eq.~(\ref{actregular}), vanishes as $\epsilon$ increases. The reason for this is straightforward: as the COV increases, there are more and more high-degree hubs. Once these recover, their neighbors'  infection rate decreases which effectively decreases $R_0$, making extinction more likely. 

For weak heterogeneity we also compared our numerical results in Fig.~\ref{fig6} to Eq.~(\ref{eq:mj_action_correction}), obtained for weakly-heterogeneous networks~\cite{miki-jason.123.068301}.  Notably, given a weakly-heterogeneous network of size $N$, the  MTE is solely determined by the mean and COV of the degree distribution $p(k)$, as the results of all networks coincide there~\cite{miki-jason.123.068301}. Yet, as $\epsilon$ increases, the MTEs of different networks depart from each other, as they are no longer independent on the higher moments of $p(k)$, see inset of Fig.~\ref{fig6}.

\begin{figure}[ht]
    \includegraphics[width=0.8\linewidth]{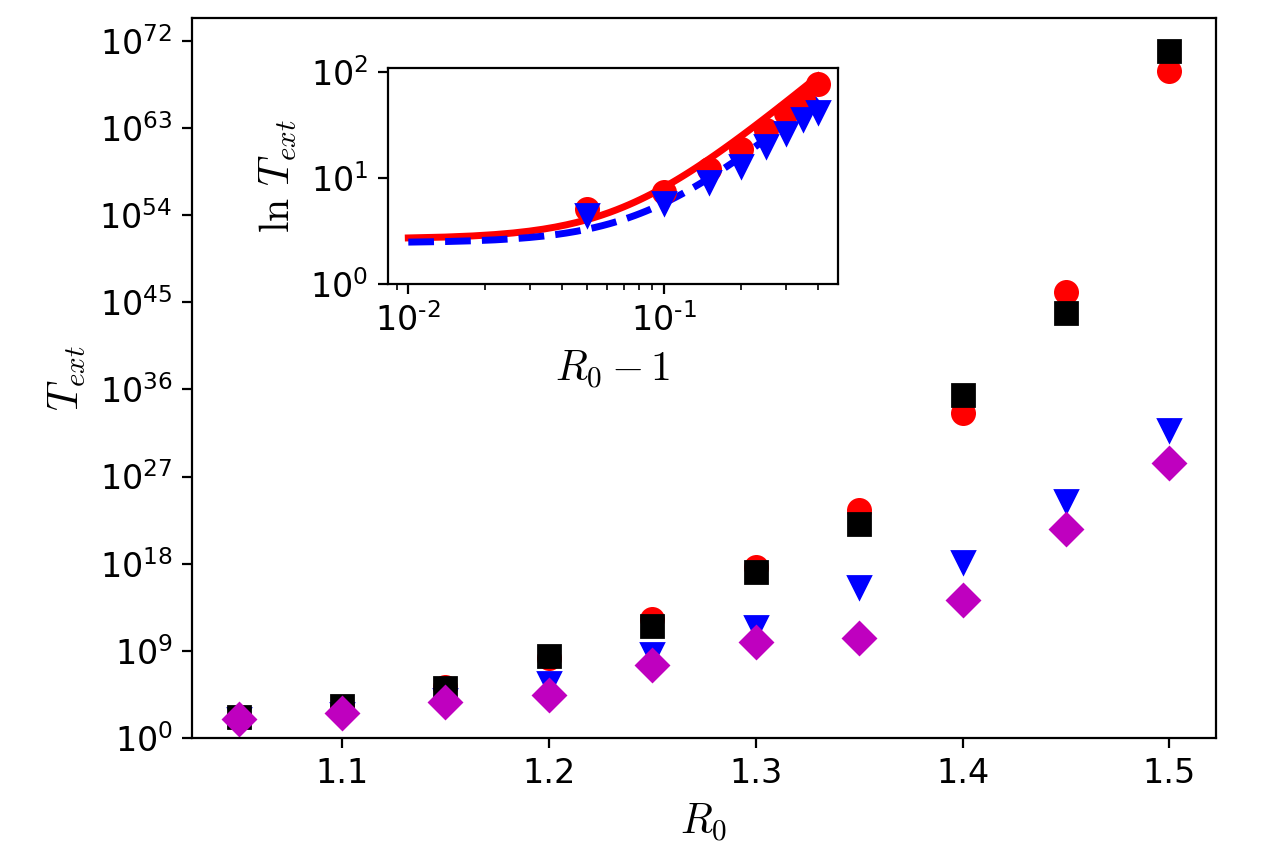}%
    \vspace{-5mm}    
    \caption{The MTE versus $R_{0}$  for  gamma (circles), inverse-Gaussian (triangles), beta (squares) and log-normal (diamonds) networks. Here $N=10^{4}$, $\left<k\right>=20$,  $\epsilon=1.5$, {\color{black}and  WE parameters $m=10^3$, $\tau_{WE}=1$ and $M=70$}.  Inset shows a log-log plot of $\ln T_{\text{ext}}$ versus $R_{0}-1$ for gamma (circles) and inverse-Gaussian (triangles) networks. The solid and dashed lines depict the analytical prediction~(\ref{eq:action_small_r}) for these networks.} 
    \label{fig7}
\end{figure}

While the network size and COV play a crucial role in determining the MTE, other parameters may also be key in determining the extinction dynamics. Such is the basic reproduction number $R_{0}$.  In Fig.~\ref{fig7}, the natural logarithm of the MTE is plotted as a function of $R_0$ for different networks with a fixed mean degree and COV. As expected, the figure shows a dramatic increase in the MTE as $R_0$ increases, with a huge variability across networks for larger values of $R_0$. 
The inset of Fig.~\ref{fig7} shows the MTEs close to bifurcation, $R_{0}-1\ll 1$, where our results agree well with Eq.~\eqref{eq:action_small_r}, obtained for arbitrary heterogeneous networks close to bifurcation~\cite{jason-ira-paths}. 

\begin{figure}[t]
    \includegraphics[width=0.8\linewidth]{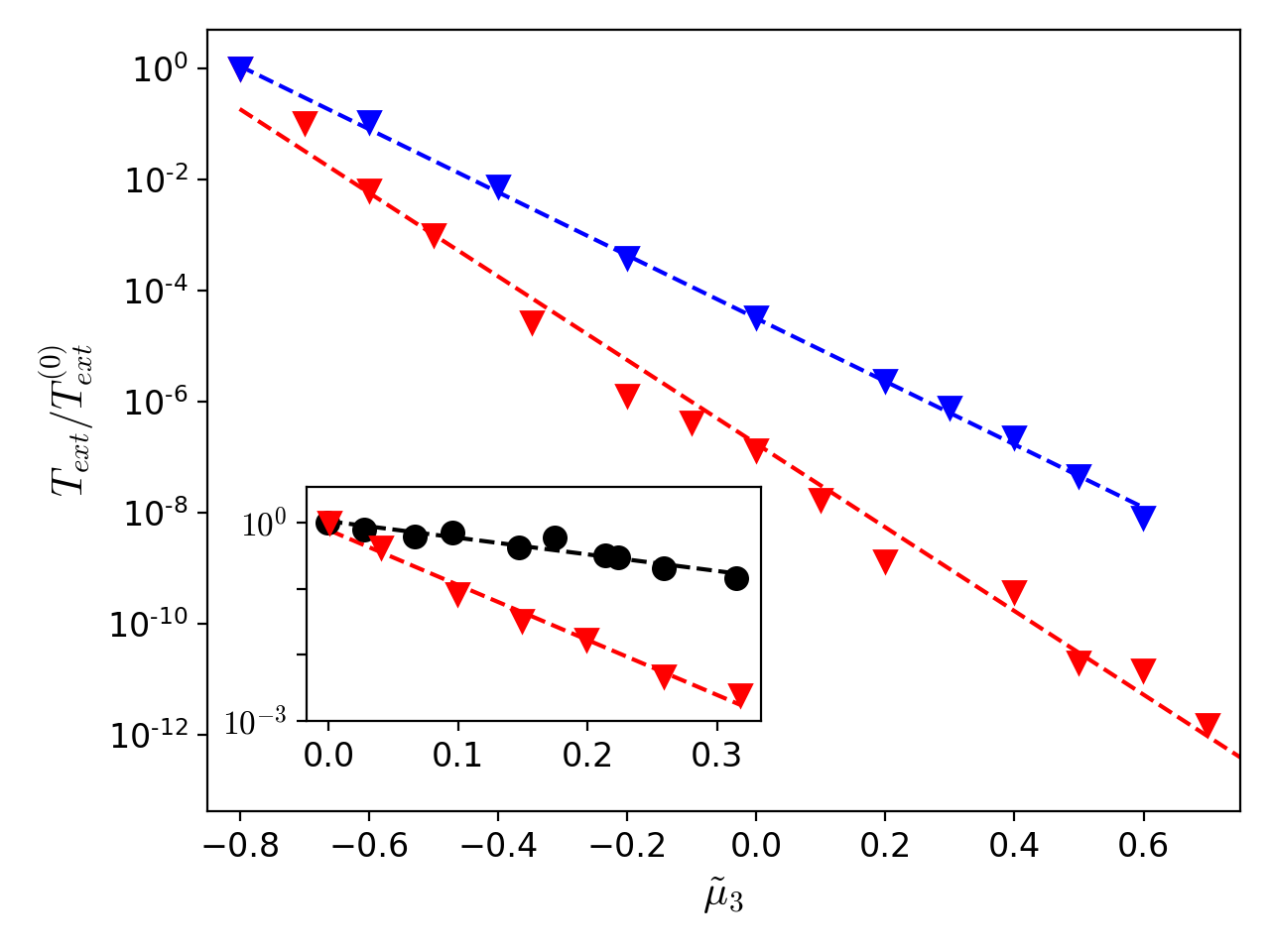}%
    \vspace{-5mm}    
    \caption{{\color{black} MTE normalized by its maximal value, $T_{\text{ext}}^{(0)}$, versus skewness $\tilde{\mu}_3$ for a network with $N=10^4$ nodes, with degree distribution given by a mixture of two Gaussians, see text. Here $R_0=1.2$ and we fix the mean \(\langle k \rangle = 50\) and the COV $\epsilon=0.6$ (red triangles) and $\epsilon=0.5$ (black triangles). The WE parameters are $m=500$, \(\tau_{WE} = 0.3\), and \( M = 70\). Inset shows the MTE normalized by its maximal value versus $\tilde{\mu}_3$ for \( N = 10^4 \), \( R_0 = 1.2\), and \(\epsilon = 0.6\) using WE parameters $m=500$, $\tau_{WE}=0.5$ and $M=70$. Simulations with fixed mean \(\langle k \rangle = 50\) and varying median (triangles), are compared with simulations with a fixed median of $50$ and varying \(\langle k \rangle\) (circles). All dashed lines represent a linear fit.}} 
    \label{fig8}
\end{figure}

{\color{black}Importantly, in Fig.~\ref{fig7}, despite the fact that we fixed the network's first two moments and size, varying $R_0$ results in significant variability in the MTE for different network distributions. The same phenomenon occurs in Fig.~\ref{fig6}, see inset. The reason is that when the COV is large, $\epsilon\gtrsim 1$, the degree distribution is highly skewed and thus, higher moments become important in determining the MTE. 
To test this, we plot in Fig.~\ref{fig8} the MTE versus the  skewness $\tilde{\mu}_3$---the third standardized moment of the distribution. Here, for simplicity, the degree distribution is given by a mixture of two Gaussian distributions with different means and equal variance, allowing us to adjust the distribution's skewness by controlling the relative heights and spread of each peak. As the skewness $\tilde{\mu}_3$ increases from a large negative value to a large positive value, the median degree moves to the left, i.e., becomes smaller. As shown in Fig.~\ref{fig8}, the MTE decreases with increasing $\tilde{\mu}_3$, despite the fact the distribution's mean is fixed. This indicates that the MTE decreases with decreasing median, which is a measure for the degree distribution's \textit{typical} degree. To further explore this dependence, in the inset of Fig.~\ref{fig8} we compare the MTEs of a calculation with varying median and fixed mean, to a calculation with varying mean and fixed median. One can clearly see that the dependence of the MTE on the median is much more pronounced than on the mean. }

{\color{black}The reason for the strong dependence of the MTE on the typical degree is as follows. In heterogeneous networks, the effective reproductive number is largely determined by the hubs, which tend to get infected and also recover quite rapidly. Once a hub recovers, the infection rate of its neighbors drops, thereby effectively decreasing $R_0$ as well as the overall MTE. As the skewness increases while keeping the mean and COV fixed, the number of hubs increase while the typical degree decreases. Therefore, a decrease in the median here is a precursor for an increase in the number of hubs, and decrease in the MTE.}

\section{Discussion \label{sec:discussion}}
We have studied the long-time dynamics and disease extinction in the realm of the SIS model of epidemics using the so-called weighted-ensemble (WE) method. In most previous works, the SIS dynamics on population networks has been analyzed using the ``brute-force" Gillespie kinetic Monte-Carlo (KMC) algorithm. This algorithm is highly efficient at short time scales, allowing to study, e.g., the epidemic threshold or stability of the endemic state on very large population networks. However, when one is interested to compute long-time dynamics such as disease clearance on large networks, the KMC algorithm becomes  infeasible, with a computation time growing exponentially with the network size. Furthermore, as heterogeneity increases, the large variability between node degrees makes standard KMC algorithms less efficient, as time steps may become vanishingly small.

These reasons emphasize the need for reliable and fast numerical schemes, such as the WE method which we have introduced here, to efficiently study rare events on heterogeneous population networks. In the WE method, as the simulation progresses, replicas that are closer to extinction are allocated more computational resources, while others are discarded, thereby enabling us to accurately compute probabilities  that were previously beyond reach with KMC methods. {\color{black}Thus, the WE method allows, e.g., computing the quasi-stationary distribution in the long-lived endemic state including the exponentially small tails, and the mean time to extinction (MTE), by measuring the total flux into the extinction state.} 

After demonstrating an equivalence between the KMC and WE methods and showing that the runtime of the latter scales linearly with the system's size, we studied the dependence of the MTE on three main features: the network's size $N$, its coefficient of variation (COV) $\epsilon$, and the basic reproduction number $R_0$. 
We have found that the dependence on $N$ in strongly heterogeneous networks is qualitatively similar to that of well-mixed systems. We have also shown that increasing heterogeneity drastically decreases the disease lifetime, whereas increasing $R_0$ strongly increases the MTE. {\color{black}An important feature that we have revealed is that even if the network's mean and COV are fixed, there can still be substantial variability in the MTE  across different networks.   To explain this, we analyzed the impact of degree skewness on the MTE and found that higher skewness leads to shorter disease lifetimes, suggesting that the median (or typical) degree serves as a more reliable indicator of MTE variation than the average degree.} Furthermore, we have managed to asymptotically corroborate previous theoretical results, which were obtained for weakly-heterogeneous networks, and when the dynamics is close to bifurcation, in parameter regimes previously inaccessible to simulations.

While we have focused on random networks, with negligible degree-degree correlations between nearest neighbors, the WE method can also be used to study assortative networks. These are characterized by the tendency of high-degree nodes  to be connected to other high-degree nodes, and vice versa; as a result they tend to percolate more easily, forming a giant component that contains a significant fraction of the entire network's nodes~\cite{newman_asso_prl,newman_asso_pre}. It has been recently shown that correlations between neighboring nodes affect the final outbreak size~\cite{leibenzon2024heterogeneity} and also the epidemic threshold; the latter, in general, is proportional to the largest eigenvalue of the adjacency matrix~\cite{anneald_net,hindes_Schwartz_2017_Epidemic_extinction_paths_in_complex_networks,Pastor-Satorras_correlated_networks,sis_degree-correlated_networks}. Yet, the study of rare events on assortative networks requires large network sizes, to overcome finite-size effects. Thus, even if the epidemic threshold can only be found numerically, the WE method is ideal for studying the interplay between network heterogeneity and assortativity. Since assortativity is  key  in many real-life social, technological, and biological networks \cite{newman_asso_prl,newman_asso_pre,anneald_net}, the WE method can thus play a crucial role in the analysis of such complex, realistic networks.

\section{Acknowledgments}
\vspace{-3mm}
\noindent
EK and MA acknowledge support from ISF grant 531/20.


\setcounter{figure}{0}
\setcounter{equation}{0}

\renewcommand{\thefigure}{S\arabic{figure}}
\renewcommand{\theequation}{A\arabic{equation}}

\section*{Appendix: Degree distributions}\label{sec:Appendix}
\vspace{-0.2cm}Here, we outline the definitions of the degree distributions employed in generating networks via the configuration model: gamma, beta, inverse-Gaussian (Wald), and log-normal distribution; see Figs.~\ref{fig1} and \ref{figS1}. For each we specify the probability density function (PDF) along with its defining parameters, which are then tied to the mean degree, $\left<k\right>$, and coefficient of variation, $\epsilon$. 

\subsubsection{Gamma-distribution}
\vspace{-0.2cm}The gamma PDF with a shape parameter $\kappa>0$ and scale parameter $\theta>0$ is given by:
\begin{equation}
    p(x;\kappa,\theta) = \frac{1}{\Gamma(\kappa) \theta^{\kappa}} x^{\kappa-1} e^{-\frac{x}{\theta}}, \quad  x \geq 0,
\end{equation}
where $\Gamma(\kappa)$ is the gamma function. Demanding that the mean and standard deviation of the gamma distribution are $\langle k\rangle$ and $\epsilon\langle k \rangle$, respectively, one has:
\begin{equation}
\hspace{-2mm}\theta = \epsilon^2 \left<k\right>, \quad \kappa =1/\epsilon^2\;\Longleftrightarrow \;\langle k \rangle=\kappa\theta,\quad \epsilon=\kappa^{-1/2}\!.
\end{equation}
Here, to obtain a fat-tailed distribution,  one must have $\kappa\lesssim 1$. If one is interested in having both large mean and large COV, one has to choose $\theta^{-1}\ll \kappa\lesssim 1$, which yields distributions such as shown in blue in Fig.~\ref{figS1}.

\subsubsection{Beta-distribution}
\begin{figure}[ht]
    \includegraphics[width=0.75\linewidth]{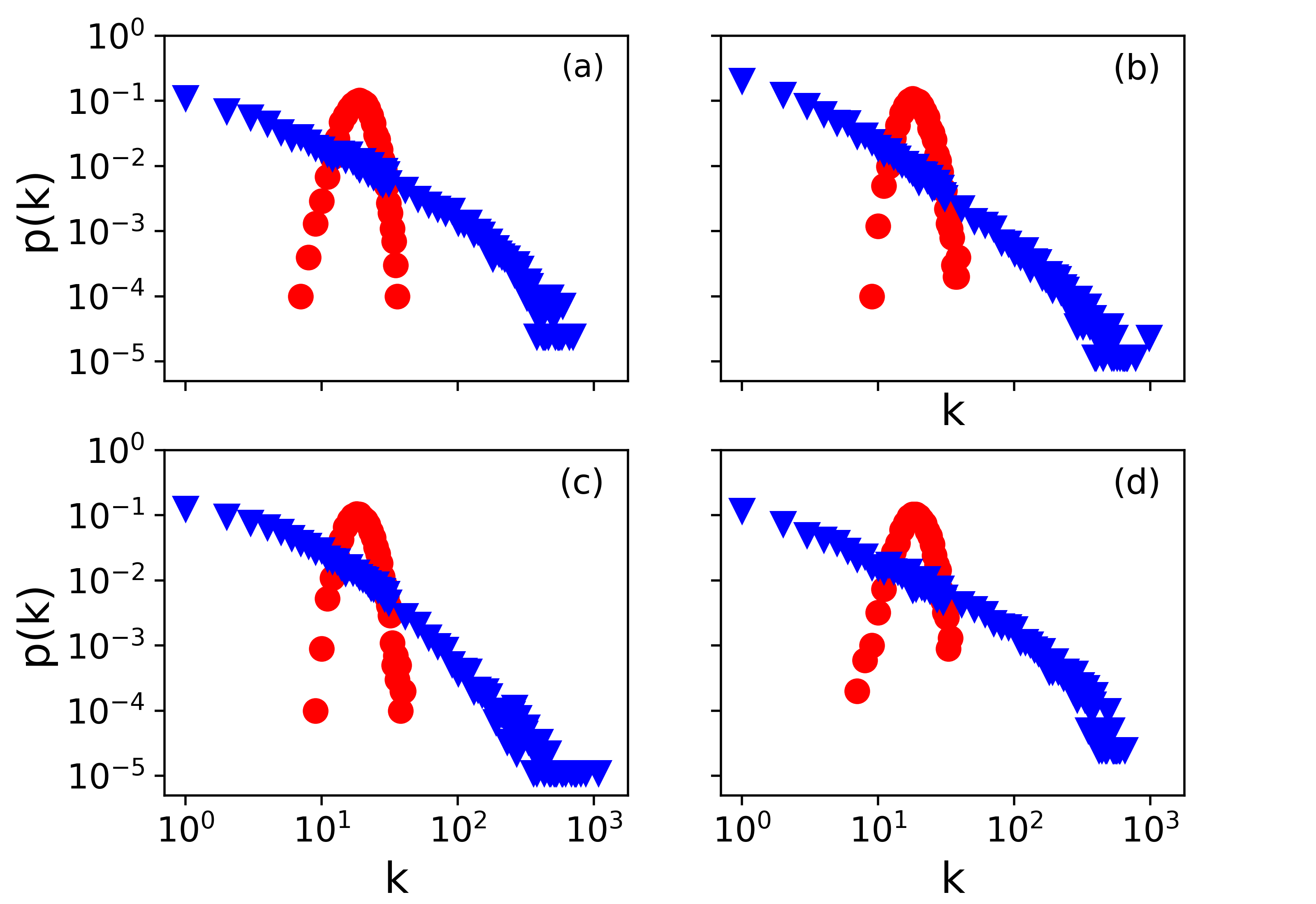}
\vspace{-4mm}    
\caption{Log-Log plot of the various degree distributions we have used,  for networks of size $N=10^{4}$ and an average degree of $\left<k\right>=20$. The distributions are Gamma (Panel a), Beta (panel b), log-normal (panel c), and inverse-Gaussian (panel d). In each panel circles denote networks with $\epsilon=0.2$, while triangles denote networks with $\epsilon=3.0$. } 
    \label{figS1}
\end{figure}

\vspace{-0.2cm}The beta PDF is given by:
\begin{equation}
    p(x; \alpha, \beta) = B(\alpha, \beta)^{-1}  x^{\alpha - 1}  (1\! - \!x)^{\beta - 1}\!, \quad 0 \leq x \leq 1.
\end{equation}
Here, $\alpha$ and $\beta$ denote the shape parameters, and $B(\alpha, \beta)$ is the beta function. Since the beta distribution has compact support, we demand that the mean and standard deviation of the beta distribution are $\langle k\rangle/N$ and $\epsilon\langle k \rangle$, respectively. In the limit of $N\gg \langle k \rangle$ we find:
\begin{equation}
    \hspace{-2mm}\alpha\simeq \frac{1}{\epsilon^2},\;\; \beta\simeq\frac{N}{\left<k\right>\epsilon^2}\;\;\Longleftrightarrow \;\;\langle k \rangle\simeq\frac{N\alpha}{\beta},\;\; \epsilon\simeq\frac{1}{\sqrt{\alpha}}.
\end{equation}
In this case, to get a fat-tailed distribution, one must have  $\alpha\lesssim 1$. If one is interested in having both large mean and large COV, one has to choose $\beta/N\ll \alpha\lesssim 1$, which yields distributions such as shown in blue in Fig.~\ref{figS1}.
\vspace{-3mm}

\vspace{-7mm}

\subsubsection{Wald distribution}
\vspace{-0.2cm}The inverse Gaussian (or Wald) PDF reads:
\begin{equation}
    p(x; \mu, \lambda) = \sqrt{\frac{\lambda}{2\pi x^3}} \exp \left[ -\frac{\lambda(x-\mu)^2}{2\mu^2x} \right], \quad x > 0.
\end{equation}
Here $\mu\!>\!0$ is the mean and $\lambda\!>\!0$ is the shape parameter. This distribution gives the time distribution of a Brownian particle with positive drift to reach a fixed positive level. Comparing with the network distribution, we find:
\begin{equation}
\hspace{-2mm}\mu=\left<k\right>,\quad \lambda = \left<k\right>/\epsilon^2\;\Longleftrightarrow \;\langle k \rangle=\mu,\quad \epsilon=\sqrt{\mu/\lambda}.
\end{equation}
Here, to obtain a fat-tailed distribution and a large mean, one has to take a large $\mu$ value and $\lambda\lesssim\mu$, which yields distributions such as shown in blue in Fig.~\ref{figS1}.

\vspace{-4mm}
\subsubsection{Log-normal distribution}
\vspace{-0.2cm}
Lastly, the log-normal PDF reads:
\vspace{-3mm}
 \begin{equation}
     p(x; \mu, \sigma) = \left(x \sigma \sqrt{2\pi}\right)^{\!-\!1} \exp\left[\!-\!\left(\ln x \!-\! \mu\right)^2\!/(2\sigma^2)\right]\!, \;\;x \!>\! 0,
 \end{equation}
where $\mu$ represents the mean of the natural logarithm of the random variable $x$ and $\sigma$ is its standard deviation. Comparing with the network distribution, we find:
\begin{eqnarray}
\mu &=& -(1/2) \ln \left[ (1 + \epsilon^2)/\left<k\right>^2 \right], \quad\sigma^2 = \ln(1 + \epsilon^2)\;\;\Longleftrightarrow \nonumber\\
\langle k \rangle&=&e^{\mu+\frac{\sigma^2}{2}},\quad \epsilon=\sqrt{e^{\sigma^2}-1}.
\end{eqnarray}
Here, to obtain a fat-tailed distribution with a large mean, one must have $\sigma\gtrsim 1$ and a large enough $\mu$, such that $\mu+\sigma^2/2\gg 1$, see distributions in blue in Fig.~\ref{figS1}.

\bibliography{bibfile}

\begin{thebibliography}{57}
\bibitem [{cod()}]{code}%
  \BibitemOpen
  \href@noop {} {\bibinfo {title} {Our codes for network generation and $\text{WE}$ numerical simulations were carried out using python and c++, and are available upon request}}\BibitemShut {NoStop}%
\end{thebibliography}

\end{document}